\documentstyle[epsfig,cite]{mn2e}
\begin{document}

\title[A highly polarised radio jet in XTE J1748$-$288]
      {A highly polarised radio jet during the 1998 outburst of the black hole transient XTE~J1748$-$288}
\author[Brocksopp et al.]
    {C.~Brocksopp$^1$\thanks{email: cb4@mssl.ucl.ac.uk}, J.C.A.~Miller-Jones$^2$, R.P.~Fender$^3$, B.W.~Stappers$^{2,4}$\\
$^1$Mullard Space Science Laboratory, University College London, Holmbury St. Mary, Dorking, Surrey RH5 6NT, UK\\
$^2$Astronomical Institute ``Anton Pannekoek'', University of Amsterdam Kruislaan 403, 1098 SJ Amsterdam, The Netherlands\\
$^3$School of Physics and Astronomy, University of Southampton, Southampton, Hants SO17 1BJ, UK\\
$^4$Stichting ASTRON, Postbus 2, 7990 AA Dwingeloo, The Netherlands\\
}
\date{Accepted ??. Received ??}
\pagerange{\pageref{firstpage}--\pageref{lastpage}}
\pubyear{??}
\maketitle
\begin{abstract}
XTE J1748-288 is a black hole X-ray transient which went into outburst in 1998 June. The
X-ray lightcurves showed canonical morphologies, with minor variations on the ``Fast
Rise Exponential Decay'' profile. The radio source, however, reached an unusually high
flux density of over 600 mJy. This high radio flux was accompanied by an exceptional ($>20\%$)
fractional linear polarisation, the variability of which was anti-correlated with the
flux density. We use this variability to discuss possible depolarisation mechanisms and
to predict the underlying behaviour of the (unresolved) core/jet components.
\end{abstract}

\section{Introduction}

Black hole X-ray transients in outburst are now well-known for exhibiting astrophysical jets, which produce synchrotron emission at radio and higher frequencies (Fender 2006 and references therein). Typical lightcurves show an initial phase during which the radio source has the flat spectrum associated with a compact jet. Later during the outburst, depending on the X-ray spectral behaviour, the radio emission may be quenched or there may be a sequence of one or more optically thin ejection events, when the radio lightcurve rises and falls with each ejection (e.g. Brocksopp et al. 2002). Finally there is typically an additional phase associated with a flat-spectrum, compact jet at the end of the outburst. Each of these phases of the radio lightcurve is connected to the underlying X-ray spectral state; in particular the behaviour of the radio source appears to be linked with the power-law component of the X-ray emission (Fender 2006 and references therein).

\begin{table*}
\caption{List of the radio observations obtained using the Australia Telescope Compact Array. }
\begin{tabular}{lccccc}
\hline
\hline
Date&MJD&\multicolumn{4}{c}{Flux Density (mJy)}\\
(1998)    &   &(1.384 GHz)&(2.496 GHz)&(4.800 GHz)&(8.640 GHz)\\
\hline
June 10 & 50974 & --      &   --        &  $82\pm   1$ &   $62\pm   1$\\
June 13 & 50977 & $433\pm 3$ &   $355\pm 4 $ &   $311\pm  2$ &   $228\pm  2$\\
June 15 & 50979 & $613\pm 3$ &   $572\pm  7 $ &  $483\pm  2$ &   $376\pm  2$\\
June 16 & 50980 & $545\pm 4$ &   $548\pm  5 $ &  $442\pm  1$ &   $328\pm  2$\\
June 22 & 50986 & $318\pm 7$ &   $302\pm  5 $ &  $238\pm  1$ &   $164\pm  1$\\
July 02 & 50996 & --      &   --        &  $66\pm   1$ &   $38 \pm  1$\\ 
July 23 & 51017 & --      &   --        &  $96\pm   1$ &   $66 \pm  1$\\ 
\hline 
\label{tab:obs}
\end{tabular}
\end{table*}

Only a few Galactic X-ray transients have been detected with a significant level of linear polarisation (LP). The first was Cyg X-3, detected at 1--5\% during a bright radio outburst soon after discovery of the radio counterpart (Gregory et al. 1972). The level of polarisation increased as the flux decayed, suggesting the evolution of a synchrotron source from optically thick to thin. Seaquist et al. (1974) later detected the source with $\sim14$\% polarisation. SS433 was discovered a few years later and found to have a radio source with 8--20\% polarisation (Hjellming \& Johnston 1981).

More recently, GRS~1915+105 was detected at 2\% LP when the emission was dominated by the approaching component as opposed to the stationary core (Rodr\'{i}guez et al. 1995). Further observations by Fender et al. (1999) detected higher, variable levels of linear polarisation (LP; 6--14\%) and highly variable position angle, again in the approaching component and not in the core. It was suggested that the variability may be due to increasing randomisation of the magnetic field within the ejecta. Additional observations of GRS~1915+105 revealed a ``linear polarisation rotator event'', when the position angle of the electric vector rotated smoothly by $50^{\circ}$ at both 4.80 and 8.64 GHz (Fender et al. 2002). With no frequency dependence, this event seemed to indicate rotation of the jet/field structure or progressive formation of a shock in the outflow. The highest fractional LP detected in GRS~1915+105 was $\sim24\%$ (Miller-Jones et al. 2005).

Hjellming et al. (1999) detected linearly polarised radio emission from the recurrent transient 4U~1630$-$47 during its 1998 outburst. This source was found to be 26\% and 18\% polarised, at 4.80 and 8.64 GHz respectively, at the peak of the radio outburst. Finally GRO~J1655$-$40 also exhibited strong linear polarisation (up to $\sim 11$\%; Hannikainen et al. 2000) during its 1994 outburst. The LP lightcurve showed complex variability which appeared inconsistent with a simple synchrotron ``bubble'' model. Instead the authors concluded that the core emission was produced in a hybrid thermal/non-thermal plasma in order to explain the Faraday rotation observed in this source. For a more detailed review of polarisation properties of X-ray binaries, see Fender (2003).

In the remainder of this section we summarise the literature published for XTE~J1748$-$288. In Sections 2 and 3 we present the results of radio monitoring of the 1998 outburst. We then discuss our results in Section 4 and present our conclusions in Section 5.

\subsection{XTE~J1748$-$288}
XTE~J1748$-$288 was discovered on 1998 June 4 using the All Sky Monitor (ASM) on-board the Rossi X-ray Timing Explorer ({\sl RXTE}) satellite (Smith, Levine \& Wood 1998). The source was also detected by the Burst and Transient Source Experiment (BATSE) on-board the Compton Gamma Ray Observatory ({\sl CGRO}), suggesting that the rise began on June 3 (Harmon et al. 1998). The X-ray source was initially hard but began to soften at BATSE energies (20--100 keV) within the first two days of monitoring. The X-ray source continued to brighten; by June 6 the X-ray colours and timing properties were still consistent with those expected for a source in the low/hard state, although the timing properties were also reminiscent of GX~339$-$4 in the very high state (Fox \& Lewin 1998).

Revnivtsev, Trudolyubov \& Borozdin (2000) made a more detailed study of the X-ray {\sl RXTE} data using both ASM and PCA (Proportional Counter Array). Based on X-ray spectral and timing analysis and comparison with GRS 1124$-$683, these authors suggested that the outburst began in the very high state but that there was an unusually dominant power-law component present. As the X-ray source faded, it passed through the high/soft and low/hard states, the 15--30 keV flux rising during the latter. Detection of an iron emission line was reported by Miller et al. (2001) and Kotani et al. (2000) using data from {\sl RXTE} and {\sl ASCA} (Advanced Satellite for Cosmology and Astrophysics) respectively.

An optically thin radio counterpart was discovered by Hjellming, Rupen \& Mioduszewski (1998a) on 1998 June 7. By June 10 the flux had risen, confirming its association with the X-ray source (Hjellming et al. 1998b; Fender et al. 1998). Finally on June 14--15, the source was resolved by the VLA, indicating proper motions of 20--40 mas per day (Rupen, Hjellming \& Mioduszewski 1998). Follow-up work by Hjellming et al. (1998c) revealed a jet velocity of $> 0.93c$ -- making it the third known Galactic source displaying apparent superluminal motion --  a distance of $> 8$ kpc and the apparent deceleration of the jet as it collided with the ISM.

\begin{figure*}
\begin{center}
\leavevmode
\epsfig{file=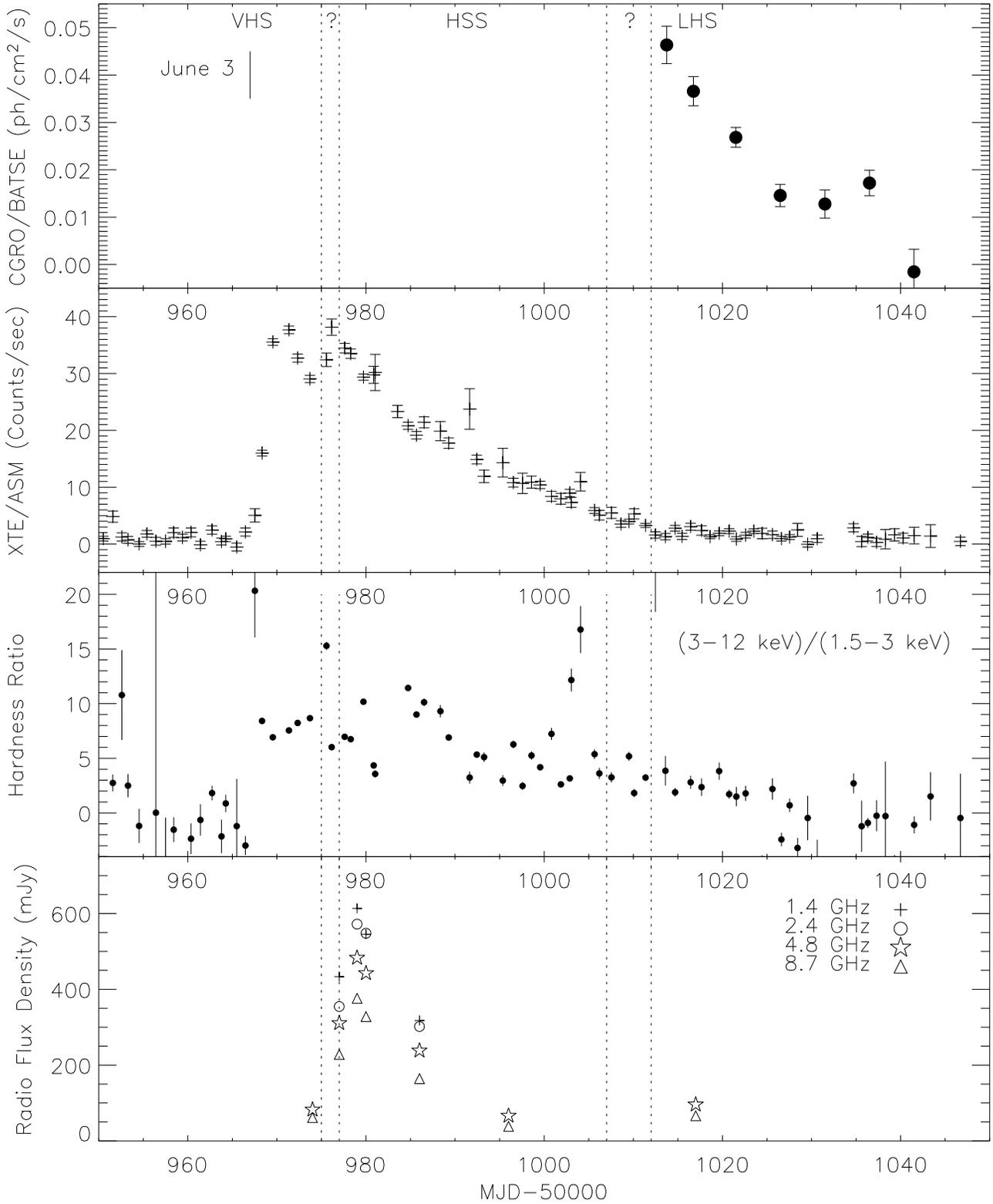, width=17cm}
\caption{X-ray and radio lightcurves of the 1998 outburst of XTE~J1748$-$288. Top: {\sl CGRO}/BATSE (20--100 keV). Second: {\sl RXTE}/ASM (1.5--12 keV). Third: {\sl RXTE}/ASM hardness ratio (3--12 keV)/(1.5--3 keV). Bottom: radio lightcurve from ATCA (with error bars smaller than the symbols). The two pairs of vertical dotted lines indicate the time-ranges during which the transitions from the very high (VHS) to high/soft (HSS) and high/soft to low/hard (LHS) X-ray spectral states are thought to have occurred (using Revnivtsev et al. 2000). The start date of the outburst, June 3 (MJD 50967), is indicated in the top panel.}
\label{fig:lightcurves}
\end{center}
\end{figure*}

\section{Observations}

\subsection{Radio: Australia Telescope Compact Array}

The Australia Telescope Compact Array (ATCA) obtained observations of XTE~J1748$-$288 on seven days during 1998 June and July, while the X-ray source was in outburst. The source was observed at 4.800 GHz and 8.640 GHz at all seven epochs, plus 1.384 and 2.496 GHz on four occasions. The array was in the 750E configuration with a bandwidth of 128 MHz. The flux and polarisation calibrator was PKS~1934$-$638 and the phase calibrator was one of PKS~B1730-130 or PKS~B1921-293, depending on the epoch.

The data were reduced using standard flagging, calibration and imaging routines from the {\sc miriad} package. Due to contamination by emission from a nearby supernova remnant, it was necessary to reject the short baselines at all wavelengths; we excluded baselines at $uv$ radii smaller than 3, 5, 10 and 20 $k\lambda$ at 1.4, 2.4, 4.8 and 8.7 GHz respectively. The {\sc miriad} routine, {\sc gpcal}, was used to solve for both instrumental and source polarisation parameters. Thus images were obtained (using natural weighting) in I, Q and U Stokes parameters, from which LP and fractional LP images could be constructed. Flux densities were obtained from the Stokes I images and are listed in Table 1.  We note that it was not possible to obtain information regarding circular polarisation; with the linear feeds of ATCA, unless the leakage terms can be determined very accurately, intrinsic circular polarisation cannot be distinguished from a small leakage of total intensity into the circular polarisation signal (Sault, Killeen \& Kesteven 1991).


\subsection{Radio: Very Large Array}

Additional radio images were obtained from the archives of the Very Large Array for comparison with the ATCA data. Data were taken at 1.425, 4.86, 8.46, 14.94, and 22.46 GHz.  The array was in its BnA configuration prior to 1998 July 3, after which it moved to B configuration. Calibration and imaging were performed using standard procedures within the National Radio Astronomy Observatory's {\sc Astronomical Image Processing System (aips)}. The flux calibrators were 3C48 and 3C286 (whichever was visible at the time of the observation). Two phase calibrators, J1744$-$3116 and J1751$-$2524, were used to interpolate the gain corrections to the target. The latter is very weak at 14.94\,GHz and could not be used at 22.46\,GHz, so was only used at the lower frequencies. Some of the observations lacked a primary calibrator, in which case the secondary calibrator was also used to set the flux density scale, averaging the flux density measured for the calibrator in previous observing runs. For observations including a primary calibrator, the secondary calibrator flux density was found to vary by of order 8

Imaging was carried out using a robust weighting scheme, biased slightly towards uniform weighting (robust parameter set to -1 or 0) in order to enhance the resolution. After a few iterations of imaging and phase-only self-calibration, the source appeared in many cases to be resolved into two components aligned approximately East-West. It was then fit with a two point source model plus a background level and slope. 

\subsection{X-ray}

X-ray monitoring data from the {\sl RXTE}/ASM amd {\sl CGRO}/BATSE instruments were also obtained from the public archives and analysed in conjunction with the radio data.

\section{Results -- radio (ATCA) and X-ray lightcurves}

\begin{figure}
\begin{center}
\leavevmode
\hspace*{-0.5cm}\epsfig{file=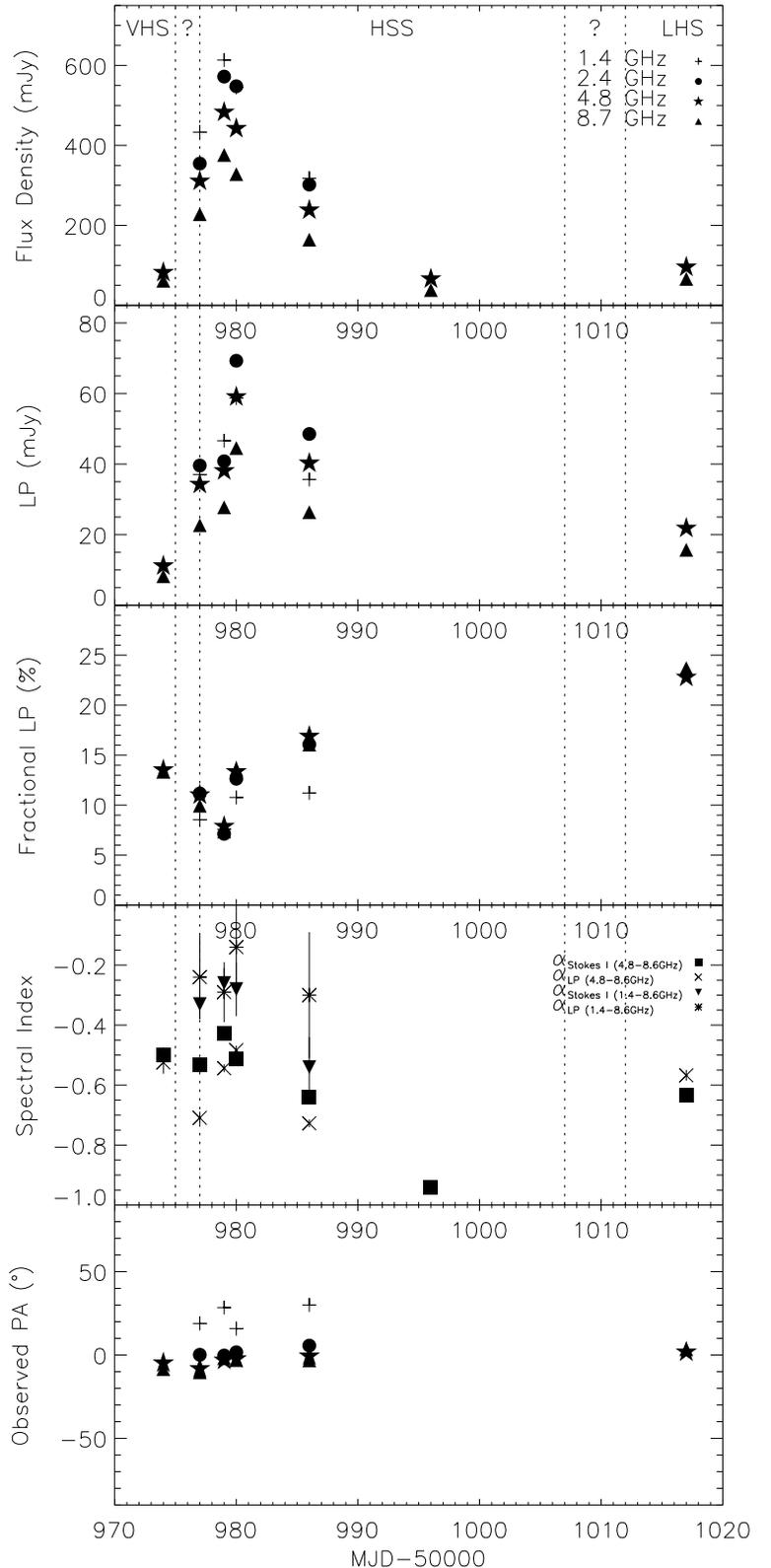, width=10cm}
\caption{Five plots showing the radio data on an expanded time axis. Top: Stokes $I$ lightcurve. Second: Linear Polarisation lightcurve ($\sqrt{Q^2+U^2}$). Third: Fractional polarisation ($100\times\sqrt{Q^2+U^2}/I$) lightcurve. Fourth: Spectral Index (obtained using both two and four frequencies and using both Stokes I and LP data-points). Bottom: Observed polarisation position angle. NB error bars are typically smaller than the symbols.}
\label{fig:radio}
\end{center}
\end{figure}

A variable point source was detected in each ATCA observation and the resultant lightcurve is plotted in the bottom panel of Fig.~\ref{fig:lightcurves}. There is a relatively steep rise to a maximum flux density of $\sim 600$ mJy at 1384 MHz, followed by a more gradual decay. The source rises again to $\sim 100$ mJy at 4800 MHz approximately a month after the decay. Compared with outbursts of other sources (e.g. GRO J1655$-$40 -- Harmon et al. 1995; XTE~J1859+226 -- Brocksopp et al. 2002) the morphology of the lightcurve is fairly straightforward. However, the peak flux density reached is unusually high, compared with the more typical values of a few mJy e.g. XTE~J1720$-$318 ($\sim 5$ mJy; Brocksopp et al. 2005), XTE~J1859+226 ($\sim 120$ mJy; Brocksopp et al. 2002).

The top panel of Fig.~\ref{fig:lightcurves} shows the available BATSE data for this outburst. It is clearly in decay by the time the observations were obtained, yet there is also a small temporary brightening before the source reaches its minimum. It is not clear whether the hard X-ray emission remained high for the duration of the soft X-ray outburst or whether this was a rebrightening as part of the same event which led to the radio rebrightening on MJD 51017. {\sl RXTE}/PCA observations by Revnivtsev, Trudolyubov \& Borozdin (2000) show a sharp dip in the 15--30 keV emission (MJD 50975 -- 51000) followed by an increase peaking on MJD 51012.

The second plot shows the ASM lightcurve. There is a fairly ``canonical'' fast-rise exponential-decay morphology but with a double-peaked maximum. The second of these two maxima is approximately coincident with the transition between the very high and high/soft spectral states, as found by Revnivtsev, Trudolyubov \& Borozdin (2000). As the soft X-ray source decays it makes the transition between high/soft and low/hard states (again, using spectral information from Revnivtsev, Trudolyubov \& Borozdin 2000) reaching the low/hard state approximately coincidentally with the onset of the BATSE observations.

The third panel in Fig.~\ref{fig:lightcurves} plots the ASM hardness ratio (3--12 keV)/(1.5--3 keV). It shows that the source was hard at the onset of the outburst and later softened. It is not clear from the ASM plots when the source returned to the low/hard state -- the transition seemed to be dominated by harder energies, as presented in Revnivtsev, Trudolyubov \& Borozdin (2000).

It is interesting to see that there is little correlation between the X-ray and radio lightcurves. The radio event appeared to begin while the X-ray source was in the very high state but most of the radio emission is observed during the high/soft state. We show in the next section that the radio emission was optically thin throughout this event, typical of ejection events as opposed to the compact jet which is associated with the low/hard state. Therefore the radio detections during the high/soft state do not contradict previous results (e.g. GX 339-4 or Cyg X-1); instead they represent discrete ejecta which are still in the process of expanding and decaying.

\section{Results -- radio polarisation}

The radio lightcurve is plotted on an expanded time axis in the top panel of Fig.~\ref{fig:radio}. While the time resolution is insufficient to be sure, it appears that there was just one dominant optically thin radio event during this period. This is supported by the fourth plot, showing the spectral index which remained optically thin at the higher frequencies at all epochs. Interestingly the spectral index during the final epoch was not zero as we would expect for the compact jet usually associated with the low/hard state. This would suggest the additional presence of residual optically thin material and we discuss the implications of this later. 

The second and third panels of Fig.~\ref{fig:radio} show the LP ($\sqrt{Q^2+U^2}$) and fractional LP (FP; $100\times\sqrt{Q^2+U^2}/I$) lightcurves respectively (but omitting the July 2 epoch due to corrupted Stokes Q and U images). The LP flux density tracks that of Stokes $I$ and reaches a maximum of $\sim$70 mJy. The FP appears to be anti-correlated with the flux density, reaching its minimum value coincidentally with the peak of the Stokes $I$ lightcurve. As the X-ray source returns to the low/hard state at the end of the outburst, the radio FP increases to its maximum value of $\sim23\%$.

\begin{figure}
\begin{center}
\leavevmode
\hspace{-1cm}\epsfig{file=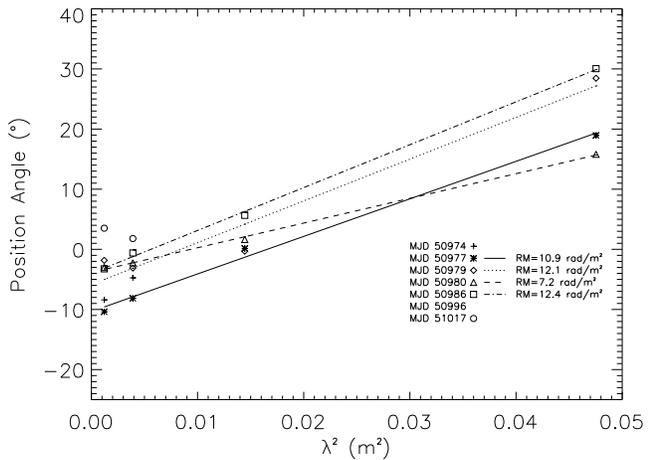, width=6cm, angle=90}
\caption{Plot showing the observed polarisation position angle plotted against $\lambda^2$ for each epoch. For the four epochs in which we have observations at four frequencies, we fit the points with a straight line in order to determine the intrinsic position angle and rotation measure. The resultant values of $RM$ are listed.}
\label{fig:pa}
\end{center}
\end{figure}

The observed polarisation position angles ($PA$) were determined from the Stokes $Q$ and $U$ images, using $PA=\frac{1}{2}\arctan(U/Q)$, and plotted in the bottom panel of Fig.~\ref{fig:radio}. In order to determine the degree to which Faraday depolarisation affects the data, we then plotted $PA$ against the square of the wavelength, according to $PA=PA_0+(RM)\lambda^2$ (Fig.~\ref{fig:pa}); this gives values for the intrinsic position angle ($PA_0$) and rotation measure ($RM$) at each epoch. This was performed twice - once for the four epochs with data at all four observing frequencies and a second time for all epochs using only the 4800 and 8640 MHz points. We find that, with a mean of 10.7 rad/m$^2$ and a range of 7.2--12.4 rad/m$^2$, the degree and variability of Faraday rotation is relatively small, compared with e.g. GRO J1655$-$40 (Hannikainen et al. 2000). We discuss the implications of this in Section 6.


\section{Results -- energetics}
The radio lightcurve obtained by ATCA is one of the few for which the rise time of a major flare can be obtained. It is therefore possible to determine the minimum power required to launch such an ejection, provided we assume that the radio source is in a state of approximate equipartition (see e.g. Longair 1994, Fender 2006). We find a lower limit to the minimum energy by assuming that the minimum energy electrons are radiating at 1.4 GHz, the lowest observing frequency. Then the minimum energy, $W_{\rm min}$ associated with an event is:

\begin{equation}
W_{\rm min}\approx3.0\times 10^6\eta^{4/7}\bigg(\frac{V}{\mbox{m}^3}\bigg)^{3/7}\bigg(\frac{\nu}{\mbox{Hz}}\bigg)^{2/7}\bigg(\frac{L_{\nu}}{\mbox{W\,Hz}^{-1}}\bigg)^{4/7}\,\mbox{J}\,
\end{equation} 

\noindent
We assume that the relativistic proton energies are negligible compared with those of the electrons, giving $\eta = 1$. The rise time of the event was 5 days; the better time-resolution of the Green Bank Interferometer (GBI; see Section 6) lightcurve confirms that there was no intervening peak. Assuming that the ejected component was spherical, we can determine a volume of $V=\frac{4\pi}{3}(ct)^3 \sim 9\times 10^{48}\, \mbox{cm}^3$. The radio lightcurve peaked at $S_0\sim 613$ mJy at $\nu_0=1.4$ GHz. Therefore the monochromatic luminosity, $L_{\nu}=4\pi d^2S_{\nu}\sim7.3\times10^{13}(d/kpc)^2\,\,\mbox{WHz}^{-1}$. Thus the minimum energy required to produce the radio ejection was 

\begin{equation}
W_{\rm min}\sim2.6(d/\mbox{kpc})^{8/7}\times 10^{35}\,\mbox{J}\,=\,2.6(d/\mbox{kpc})^{8/7}\times10^{42}\,\mbox{erg}\,
\end{equation}

\noindent
By dividing this by the rise-time (5 days), we obtain the minimum jet power:

\begin{equation}
P_{\rm min}\sim6.0(d/\mbox{kpc})^{8/7}\times 10^{29}\,\mbox{J\,s}^{-1}\,=\,6.0(d/\mbox{kpc})^{8/7}\times10^{36}\,\mbox{erg\,s}^{-1}
\end{equation}

We note that our assumed rise-time is based on the increase of flux density in the radio lightcurves. The spectral indices plotted in Fig.~\ref{fig:radio} suggest this is a reasonable assumption based on the higher frequencies, which show fairly optically thin emission ($\alpha \sim -0.5$). However, the spectral index between 1.4 and 8.7 GHz is less optically thin, suggesting possible optical depth effects at the lower frequencies and a shorter intrinsic rise time. In that case the parameterisations of Equations 2 and 3 become $W_{\rm min}\sim 3.3\,(t/\mbox{days})^{9/7}(d/\mbox{kpc})^{8/7}\times10^{41}\,\mbox{erg}$ and $P_{\rm min}\sim 3.8\,(t/\mbox{days})^{2/7}(d/\mbox{kpc})^{8/7}\times10^{35}\,\mbox{erg\,s}^{-1}$

We further note that Equation 1 for the minimum energy assumes an optically thin radio source with spectral index of $-0.75$, compared with the value $-0.42$ observed for XTE~J1748$-$288 (assuming the convention $S_{\nu}\propto\nu^{\alpha}$ where $\alpha$ is the spectral index). Adapting the standard parameterisation of Equation 1 in order to account for the spectral index, results in a small increase in minimum energy/power by a factor of 1.24

Finally, we need to account for Doppler (de-)boosting;  Adopting a jet velocity of $\sim 0.93c$ from Hjellming et al. (1998c), we obtain a possible range of Doppler factors, $\delta=\gamma(1\mp\beta\cos \theta)^{-1}\sim$ 1.4--38.6, depending on the inclination angle of the system and whether the jet is approaching or receding. Thus the scaling factors, $\gamma\delta ^{-5/7} $ and $\gamma\delta ^{-12/7} $ change the values of $W_{\rm min}$ and $P_{\rm min}$ by factors of 0.2--2.1 and $5e ^{-3}$--1.5 respectively.

The distance to XTE~J1748$-$288 is not known (although see also Hjellming et al. 1998c). The distance of the Galactic centre, however, is $\sim 8$ kpc (e.g. Eisenhauer et al. 2003) and so, given the location of the XTE~J1748$-$288 towards the Galactic centre, such a value might be considered a reasonable maximum distance, yielding $P_{\rm min}\sim 8.0\times 10^{37}\mbox{erg\,s}^{-1}$ (excluding the Doppler factor). Considering that (i) the ATCA observations were obtained over a relatively narrow wavelength range, and (ii) we do not take into account the energy associated with bulk relativistic motion, this is probably a fairly conservative estimate. Nonetheless, it is comparable with the minimum power required by one of the ejections of GRS~1915+105 (Fender et al. 1999). Clearly a sizeable fraction of the accretion power is needed to eject such a powerful jet, perhaps even more so in XTE~J1748$-$288 if we make the reasonable assumption that the mass of the black hole is less than that of GRS~1915+105; with a large estimated mass of $14\pm 4\mbox{M}_{\odot}$, GRS~1915+105 challenges theories of black hole formation in binary systems (Greiner, Cuby, McCaughrean 2001).

\section{Discussion}

On first inspection, it appears that XTE~1748$-$288 was an exception to the ``rule'' that all X-ray transients begin their outbursts in the low/hard state. However, regardless of whether this outburst originated in a brief low/hard or very high X-ray spectral state, Revnivtsev, Trudolyubov \& Borozdin (2000) report the presence of a dominant hard power-law component at low X-ray energies (akin to the ``hard'' intermediate and/or very high states proposed by Fender, Belloni \& Gallo 2004). It is this dominant power-law component which has been present at the beginning of all black hole X-ray transient outbursts and which is typically associated with jet emission (e.g. Fender 2006 and references therein). The labelling of ``canonical'' spectral states may be disadvantageous here -- consideration of whether the spectrum was disc- or jet/corona-dominated (i.e. blackbody- or power-law-dominated) may be more useful and physical (see Brocksopp et al. 2006 for further discussion).

For a uniform magnetic field, the degree of LP of an optically thin synchrotron source is determined by $(p+1)/(p+\frac{7}{3})$, where $p$ is the electron energy power index. We would therefore expect optically thin synchrotron sources, as observed during the ejection events, to be up to 60--70\% polarised. Yet, as discussed in Section 1, only a few Galactic jet sources have been detected with significant FP. 

Of course, the majority of these radio sources cannot be assumed to have uniform magnetic fields but the LP of a source can be reduced by a number of other effects; some are intrinsic, others are dependent on the location of the observer. Some radio sources might be comprised of multiple components, such as a weakly-polarised core and highly-polarised jets. This appeared to be the case for e.g. GRS~1915+105 (Fender et al. 1999) and is often seen in radio galaxies (e.g. Cawthorne et al. 1993). The observed radio emission from an unresolved source will have polarisation properties dependent on the relative contributions of the jet and core. Alternatively a source might be comprised of various polarised ``packets'', but each with diferent position angles. The net effect will be to reduce the observed polarisation, as is observed in SS433 where the rotating jets have the effect of contributing ``packets'' of linearly polarised emission with different position angles (Stirling et al. 2004). Finally the degree of polarisation may be reduced through rotation of the position angle, due to one of two mechanisms. The first is being viewed through multiple magnetic media i.e. Faraday rotation, where the degree of rotation is proportional to $\lambda^2$. The second is via a wavelength-independent mechanism, e.g. GRS~1915+105, which could be interpreted as physical rotation of the magnetic field structure, formation of a shock at an angle inclined to the line of sight, a helical magnetic field structure (as discussed in Fender et al. 2002) or acceleration/deceleration of the jet (Blandford \& K\"onigl 1979).

\begin{figure}
\begin{center}
\leavevmode
\epsfig{file=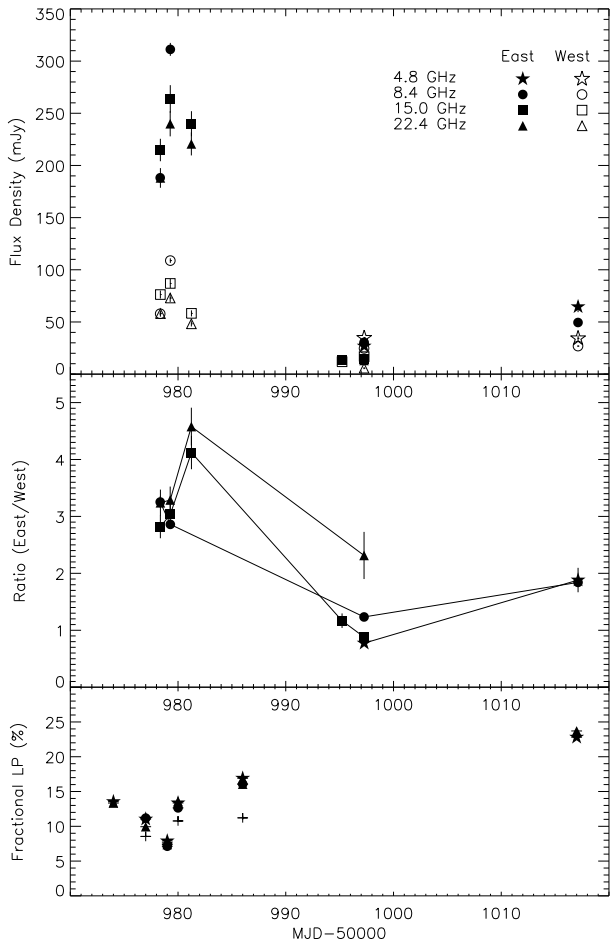, width=8cm, angle=0}
\caption{Top: Radio lightcurves from the VLA, using data from epochs most contemporaneous with the ATCA data. We plot separately the contributions from the east (filled symbols) and west (open symbols) components resolved in the images. Middle: Ratio plot of the east:west contributions. Bottom: The FP lightcurve, copied from Fig.~\ref{fig:radio} for comparison. The eastern component dominates the emission but its relative contribution decreases during the epochs when the ATCA data shows the highest fractional polarisation. This is consistent with our suggestion that the eastern component is the depolarised core and the western component the linearly polarised jet.}
\label{fig:vla}
\end{center}
\end{figure}

In the case of XTE~J1748$-$288, we cannot rule out the possibility that variable Faraday rotation is responsible for the variability in FP -- however, the rotation measure is relatively small and non-variable compared with e.g. GRO J1655$-$40 (Hannikainen et al. 2000). Alternatively, the FP lightcurve is anti-correlated with the flux densities, perhaps suggesting that the peak of the outburst is dominated by a depolarised core. As the core weakens, the relative importance of the jet component, and hence the FP, increases. 

Of course, the ATCA images were unresolved and so any discussion of core versus jet emission remains speculative. With this in mind, we have obtained archival VLA data for the epochs most contemporaneous with ATCA epochs. All of the VLA images were resolved at at least one frequency (except 1998 June 22, when the source elevation was very low; 10--11$^{\circ}$), showing two components aligned east-west with position angle $\sim86^{\circ}$ east of north . In Fig.~\ref{fig:vla} we plot the flux densities of these two components in the top panel, with the ratio of the flux densities in the middle panel and the FP lightcurve from Fig.~\ref{fig:radio} replotted for comparison. These VLA plots show that during the major ejection the emission of the eastern component was considerably brighter than that of the western component. As the major flare subsided, so too did the relative importance of the eastern component. If the eastern component can be identified with the core (and perhaps a receding jet) and the western component with the approaching jet, then the VLA data confirm our speculative explanation for the increase in FP as the jet component travelled outwards from the core. A more detailed analysis of the VLA data is beyond the scope of this paper but will be addressed in a later paper.

If such a scenario is correct then we must question what causes the high FP at the beginning of the lightcurve presented in this paper. The obvious suggestion is that it is the remnant of either an unobserved previous ejection, or of a compact jet which would have been present during an initial low/hard state phase. For a self-absorbed synchrotron source, as is associated with the low/hard state, the maximum polarisation is $3/(6p+13)=$10--15\% (Longair 1994). This is not inconsistent with the $\sim14$\% level detected, although extremely unlikely, given that the greatest previously observed polarisation of a low/hard state source is $\sim 2\%$ for GX~339$-$4 and GS~2023+338 (Corbel et al. 2000; Han \& Hjellming 1992). The high flux density at this time is also inconsistent with other systems and the ``universal'' X-ray:radio correlation of Gallo, Fender \& Pooley (2003). Therefore a model invoking optically thin radio emission may be more appropriate, such as the internal shock model of Fender, Belloni, Gallo (2004); the radio peak and FP minimum correspond to the collision of the fast jet with an earlier, slower jet which was associated with an early low/hard state jet (perhaps causing the deceleration observed by Hjellming et al. 1998c). The dip in FP would take place as the magnetic field of the ejecta became temporarily chaotic. The shocks would then compress and reorder the magnetic field, thus providing a mechanism for the return to a high level of polarisation.

\begin{figure}
\begin{center}
\leavevmode
\epsfig{file=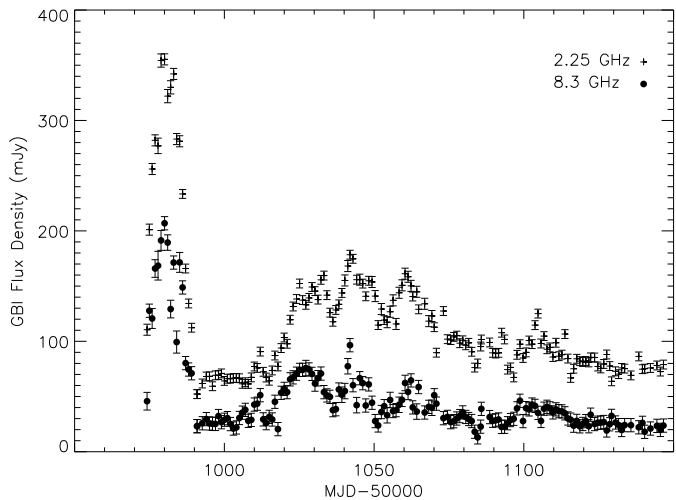, width=6.5cm, angle=90}
\caption{GBI lightcurves showing the peculiar variability of the final low/hard state. {\em Please note that we do not trust the absolute flux density values due to the discrepancy in calibration bewteen the GBI and the ATCA/VLA.}}
\label{fig:gbi}
\end{center}
\end{figure}

The FP is again high ($\sim23$\%) once the X-ray source has returned to the low/hard state and, again, we cannot attribute the high -- and rising -- radio flux density at this time to a ``normal'' low/hard state compact jet. Instead, it appears to be optically thin radio emission from some additional source. However we have also looked at public data obtained with the GBI and plot the lightcurve in Fig.~\ref{fig:gbi}. These data are not included in our analysis on account of a discrepancy between the absolute flux calibration of the GBI compared with the ATCA and VLA. Since the GBI was a single-baseline array and is now obsolete, we do not trust the absolute flux densities in this plot, however, the variability is striking. During the low/hard state, the radio source falls to a bright plateau state (MJD 50990 onwards), which is later superimposed by further episodes of flaring  (MJD 51100 onwards). The signal-to-noise and time resolution of the ASM and BATSE lightcurves do not allow us to look for correlated X-ray behaviour and there are no previous transient events with which we can compare this behaviour. This is not typical behaviour for residual optically thin emission as it fades. We therefore suggest three alternative mechanisms. The first is that it is a new sequence of ejections taking place as the X-ray source makes the transition from high/soft to low/hard states. This would contradict the model of Fender, Belloni, Gallo (2004) but is not unprecedented, given the radio flare observed in Cyg X-1 as it made the transition from soft to hard in 1996 (Zhang et al. 1997). A second suggestion is that the internal shock model applies repeatedly within a multiple-speed flow, and that these peaks are the shocked emission as previously-ejected high-Lorentz factor material again collides with the original emission of the low/hard state jet. Such a mechanism may explain the absence of any simultaneous X-ray variablity. Finally it may be that this emission is produced through interaction betwen the jet and interstellar medium, analogous to the lobes seen in Fanaroff-Riley II radio galaxies and as observed in XTE~J1550$-$564 (Corbel et al. 2002; Kaaret et al. 2003). The high density of the surrounding Galactic centre region makes this plausible, despite the general lack of success in searching for the equivalent of hotspots in Galactic jet objects (but see also Gallo et al. 2005). However, preliminary work by Brocksopp, Kaiser, Schoenmakers in prep. studies some FRII radio galaxies which also lack hotspots and this work aims to increase our understanding of how the jet energy might be dissipated in the ISM/IGM. Again, a detailed analysis is beyond the scope of this paper but further analysis of the higher-resolution VLA data may help distinguish between these alternatives.

\section{Conclusions}
We have obtained radio monitoring of the 1998 outburst of the black hole X-ray transient XTE~J1748$-$288 and compared the data with public X-ray data. While there are aspects of the outburst which follow canonical patterns, such as the lightcurve morphology and X-ray spectral behaviour, this outburst was exceptional for its radio emission. The radio source reached a peak of $\sim600$ mJy, making it one one of the brightest amongst the black hole X-ray transients. The high radio flux density was accompanied by high ($>20\%$) FP; the FP lightcurve was anti-correlated with the radio lightcurve. This might suggest that most of the polarised flux was in an extended component, while most of the remaining emission was from a depolarised core. We used VLA images to add weight to these conclusions. Alternatively, the reduction in FP could be attributed to an internal shock within the jet which temporarily disordered the magnetic field.

\section*{acknowledgments}
The Australia Telescope is funded by the commonwealth of Australia for operation as a National Facility managed by the CSIRO. The National Radio Astronomy Observatory is a facility of the National Science Foundation, operated under cooperative agreement by Associated Universities, Inc. The Green Bank Interferometer is a facility of the National Science Foundation operated by the National Radio Astronomy Observatory.  From 1978-1996, it was operated in support of USNO and NRL geodetic and astronomy programs; after 1996 in support of NASA High Energy Astrophysics programs. CB is funded by PPARC.

\end{document}